\documentclass[traditabstract, bibyear]{aa} 
\usepackage{graphicx}
\usepackage{txfonts}
\usepackage{color}
\usepackage{amstext}
\usepackage{answers}

\begin{document}

   \title{Are all RR Lyrae stars modulated?}

   \author{Geza Kovacs\inst{1}}

   \institute{Konkoly Observatory of the Hungarian Academy of Sciences, 
              Budapest, 1121 Konkoly Thege ut. 13-15, Hungary \\
              \email{kovacs@konkoly.hu}
             }

   \date{Received 6 April 2018 / Accepted xx, xx 2018}

 
  \abstract
{We analyzed $151$ variables previously classified as fundamental mode 
RR~Lyrae stars from Campaigns $01-04$ of the {\em Kepler} two wheel (K2) 
archive. By employing a method based on the application of systematics 
filtering with the aid of co-trending light curves in the presence of 
the large amplitude signal component, we searched for additional Fourier 
signals in the close neighborhood of the fundamental period. We found 
only $13$ stars without such components, yielding the highest rate of 
$91$\% of modulated (Blazhko) stars detected so far. A detection 
efficiency test suggests that this occurrence rate likely implies a 
$100$\% underlying rate. Furthermore, the same test performed on a 
subset of the Large Magellanic Cloud RR~Lyrae stars from the MACHO 
archive shows that the conjecture of high true occurrence rate fits 
well to the low observed rate derived from this database.    
}

   
\titlerunning{Are all RR Lyrae stars modulated?}
\authorrunning{Kovacs, G.}
   \maketitle
%

%
%
\section{Introduction}
Although RR~Lyrae stars are among the most important objects for 
tracing galaxy structure and chemistry and thought to pulsate in 
the simplest way we can model by and large already from the mid 
sixties (Christy~\cite{christy1964}), it is highly disturbing that 
when it goes to a more detailed modelling (e.g., steady multimode 
pulsation), our current (admittedly simple) models fail in a major 
way. Particular to these stars is the (quasi)periodic amplitude/phase 
change, commonly known as Blazhko phenomenon (Blazhko~\cite{blazhko1907}, 
see also Shapley~\cite{shapley1916}). Most of the progress made in 
this issue comes from the side of the observations, largely attributed 
to the space missions {\em CoRoT} and {\em Kepler} 
(e.g., Benk\H{o} et al.~\cite{benko2014}), Szab\'o et al.~\cite{benko2014} 
Still, even these highly accurate data were unable to deliver a 
ground-braking discovery that would inspire physically sound (and working) 
idea.\footnote{The only exception is perhaps the discovery of the alternation 
of the maxima in certain phases of the Blazhko cycle, that is currently 
attributed to a high-order resonance involving the fundamental mode 
-- see Buchler \& Kollath~(\cite{buchler2011}) and  Smolec~(\cite{smolec2016}) 
for some critical comments.} 

There are also contradictory figures on the mere occurrence rates of the 
Blazhko (hereafter BL) stars. These rates (for fundamental mode stars) 
range from $\sim 12$\% to $\sim 50$\% (see Kovacs \cite{kovacs2016} and 
Smolec~\cite{smolec2016} and and references therein). Interestingly, the 
analysis of the high-precision space observations yield apparently quite 
similar rates to those derived from classical ground-based data 
(Benk\H{o} et al.~\cite{benko2014}, Jurcsik et al.~\cite{jurcsik2009}). 

This, and the apparent lack of the application of the modern methods of 
time series analysis in this field prompted us to examine the occurrence 
rates of the BL stars from a subset of the available space data. Even 
without knowing the underlying physical cause of the BL phenomenon, it 
is clear that this single number may help to distinguish between various 
ideas. For example, models based on the resonance between radial modes 
will likely face difficulties if amplitude/phase modulation is indeed 
as common as the present study shows.

%
%
\section{Datasets and the method of analysis}
\begin{figure}
 \vspace{0pt}
 \centering
 \includegraphics[angle=-90,width=85mm]{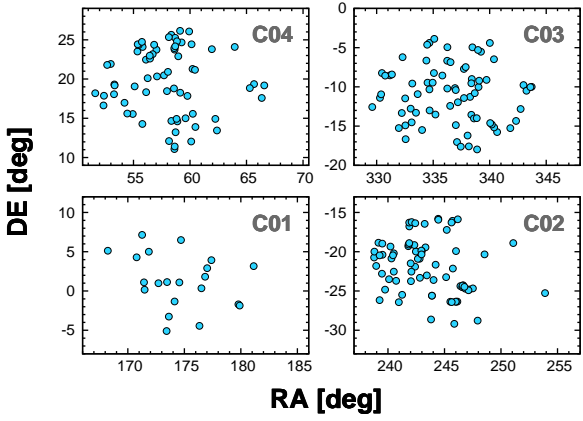}
 \caption{Distribution of the stars analyzed in this papers (see Table~\ref{datasets}) 
          in the field of view of {\em Kepler}.}
\label{ra_de}
\end{figure}

In establishing a solid, but not too extensive dataset proper for this 
introductory work, we opted for the {\em Kepler} two-wheel (K2) survey. 
Unfortunately, there are not too many well-documented (star-by-star) 
RR~Lyrae surveys available for this mission. However, the variability 
survey of Armstrong et al.~(\cite{armstrong2016}) of the data of 
Campaigns 0--4 has proven to be extensive enough, suitable for the 
purpose of this work. We have chosen not to include C0, because of the 
short observational time span devoted to this campaign in the early 
phase of K2. A few basic parameters of the datasets used are listed 
in Table~\ref{datasets}. As this table shows, there are several 
miss-classifications in the original list. Nevertheless, the number 
of the remaining stars is large enough to use them as a statistically 
significant sample in assessing the occurrence rate of the BL phenomenon. 
There are some 4--5 stars belonging to the globular cluster M4 in C02, 
otherwise all the rest are apparently Galactic field/halo stars (many 
of them are rather faraway objects). See Fig.~\ref{ra_de} for the 
distribution of the $237$ targets analyzed in this paper over the 
{\em Kepler} field of view during the four campaigns.    

%
%
\begin{table}[!h]
  \caption{Summary of the datasets used in this paper}
  \label{datasets}
  \scalebox{1.0}{
  \begin{tabular}{crrl}
  \hline
   Field    & N$_{\rm tot}$  & $N_{\rm RR}$  &  Source \\ 
 \hline
C01  &   21   & 11  & IPAC/K2 ExoFOP \\
C02  &   76   & 45  & IPAC/K2 ExoFOP \\
C03  &   73   & 62  & IPAC/K2 K2 Targets \\
C04  &   67   & 33  & IPAC/K2 K2 Targets \\ 
\hline
\end{tabular}}
\begin{flushleft}
{\bf Notes:}\\
\vbox{
N$_{\rm tot}$ stands for the total number of stars classified as RRab 
stars by Armstrong et al.~(\cite{armstrong2016}). $N_{\rm RR}$ is for 
the actual number of RRab stars. The time series data were 
downloaded for C01 and C02 from \url{https://exofop.ipac.caltech.edu/} \\ 
(see Petigura et al.~\cite{petigura2015}) and for for C03 and C04 from \\
\url{https://exoplanetarchive.ipac.caltech.edu/} \\} 
\end{flushleft}
\end{table}

For fields C01 and C02, time series based on Simple Aperture Photometry 
(SAP) have been downloaded in ASCII format from the target inquiry page 
of the NASA Exoplanet Archive (IPAC). Field C03 and C04 time series data 
are not available at this site, but they are accessible through the 
ExoFOP link by the same host. To aid the method vital for filtering out 
instrumental systematics (TFA, Kovacs et al.~\cite{kovacs2005}), in 
addition to the fundamental mode RR~Lyrae (RRab) stars, 2--4 thousand 
stars were also downloaded for each field. These stars are distributed 
over the entire {\em Kepler} field of view and cover a wide range of 
brightness. All time series are of long cadence (i.e., with $\sim 30$~min 
sampling) and contain over $3000$ data points with a continuous coverage 
of $\sim 70$~days. 

%
%

It is well known that instrumental systematics are much more severe 
in the K2 phase of the {\em Kepler} mission than it was during the first 
phase with all four reaction wheels working. Therefore, 
correction for the various forms of systematics originating from 
this, and several other issues (e.g., ignition of thrusters) should 
be considered during the time series analysis. In the case of large 
amplitude variable stars this correction (if it is made without full 
modelling of the data) may lead to disastrous result. We guess that 
this is the main reason why variability studies, so far, largely 
avoided any filtering of systematics and tried to reach the best 
result with the aid of carefully chosen aperture forms and use only 
aperture photometry without any essential post-processing 
(e.g., Plachy et al.~\cite{plachy2017}). 

To search for astronomical signals and tackle systematics at the 
same time, without deforming the signal, 
Foreman-Mackey et al.~(\cite{foreman-mackey2015}) and 
Angus et al.~(\cite{angus2016}) employed full models, whereas 
Aigrain et al.~(\cite{aigrain2016}) developed a method based on 
the flexibility of modelling with Gaussian Process. As discussed 
by Kovacs et al.~(\cite{kovacset2016}), when these methods are used 
for period search for signals commensurable with the size of the 
systematics, due to the extra freedom introduced by the inclusion 
of the underlying (but unknown) signal, the resulting detection 
statistics becomes poorer than for the more standard methods, 
assuming no signal content (e.g., SysRem of Tamuz et al.~\cite{tamuz2005}). 

The situation is different when the period of the dominating signal 
is known and we search for small secondary signals hidden in the 
systematics. The very simple methodology one can follow has been 
touched upon in some of our earlier papers (Kovacs~\cite{kovacs2005}, 
Kovacs \& Bakos~\cite{kovacs2008}). We described the method and  
illustrated its power through examples drawn from the K2 database 
in Kovacs~(\cite{kovacs2018}). For the sake of a more compact 
presentation, here we summarize the main steps of the analysis. 
\begin{itemize}
\item[1.]
Choose $N_{\rm TFA}$ co-trending time series 
$\{U_{\rm j}(i); j=1,2, ..., N_{\rm TFA}; i=1,2, ..., N\}$ 
from the stars available in the field (interpolate -- if needed -- 
to bring the target and the co-trending time series to the same timebase). 
Note: in this study, we selected these stars from the bright end of the 
available stars, but no other selections were made (e.g., on the basis 
of variability). 
\item[2.]
Derive the fundamental frequency $f_0$ of the target 
$\{T(i); i=1,2, ..., N\}$ from the SAP time series. 
\item[3.]
Compute the sine, cosine values up to the $N_{\rm FOUR}-1$ 
harmonics of $f_0$: 
$\{S_{\rm j}(i), C_{\rm j}(i); j=1,2, ..., N_{\rm FOUR}; i=1,2, ..., N\}$
\item[4.] 
Perform standard Least Squares minimization for $a_0$, 
$\{a_{\rm j}\}$ $\{b_{\rm k},c_{\rm k}\}$ on the following expression: 
\begin{eqnarray}
\label{eq:1}
\mathcal D & = & \sum_{i=1}^N [T(i) - F(i)]^2 \hskip 2mm ,\\
F(i) & = & a_0 + \sum_{j=1}^{N_{\rm TFA}}a_{\rm j} U_{\rm j}(i) + 
           \sum_{k=1}^{N_{\rm FOUR}}b_{\rm k} S_{\rm k}(i) + c_{\rm k} C_{\rm k}(i)
\hskip 1mm .
\end{eqnarray}  
\item[5.]
Using the solution above, compute 
$\hat{T(i)}=T(i) - a_0 - \sum_{j=1}^{N_{\rm TFA}}a_{\rm j} U_{\rm j}(i)$ 
and find the new value of $f_0$. Iterate on 3., 4. and 5., until convergence 
is reached. 
\item[6.]
With the converged $f_0$, compute the residual $\{R(i)=T(i)-F(i); i=1,2, ..., N\}$ 
and perform standard Fourier frequency analysis on $\{R(i)\}$ with pre-whitening. 
\item[7.]
After reaching the noise level, use all frequencies ($f_0$ and its harmonics, the 
new frequencies) and the co-trending time series to perform a grand fit 
according to step 4 (extended with the new frequencies). 
\end{itemize}   
We note that certain care must be taken at some steps of the analysis. For example, 
if the systematics are too large, then we need to perform a TFA analysis first, 
to derive a good estimate on $f_0$ (we had one such a case in the dataset analyzed). 
Also, some frequency proximity condition should be applied, otherwise the fit 
may become unstable. We found that requesting $|\Delta f|<0.1/T$, where $T$ is 
the total time span, yields stable fits.

%
%
\section{Analysis, examples, observed occurrence rate}
We analyzed all the $237$ objects of Table~\ref{datasets} by using the 
method described in Sect.~2. A wide frequency band of $[0,30]$~c/d was 
chosen to accommodate high harmonics. Even though the number of harmonics 
was high ($14$, throughout the analysis), in many cases leakage from the 
harmonics higher than the Nyquist frequency ($24.47$~c/d in the case of 
the present K2 data) caused a well-defined pattern from $\sim 10-15$~c/d. 
Our stoppage criterion for ending the pre-whitening cycle was $SNR<6$, 
where $SNR=(A_{\rm peak}-\langle A \rangle)/\sigma(A)$, with $A$ denoting 
the amplitude spectrum with sigma-clipped average $\langle A \rangle$ 
and standard deviation $\sigma(A)$. The results presented in this paper 
are based on runs using $100$ co-trending stars. We note that in the 
preliminary study we used $400$ and we ended up with the same conclusion. 
In critical cases we used different number of co-trending light curves, 
to make sure that the final classification is robust enough.  

%
%
\begin{figure}
 \vspace{0pt}
 \centering
 \includegraphics[angle=-0,width=75mm]{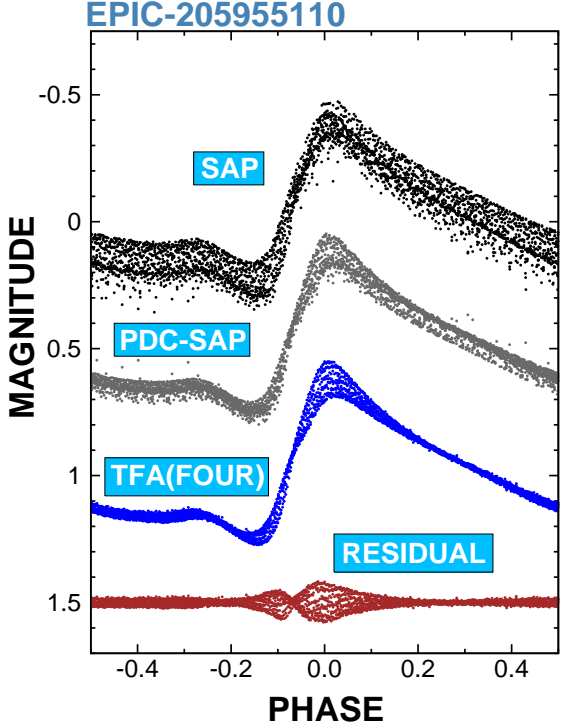}
 \caption{An example from from Campaign 3 on the performance of the filtering 
          method used in this work in searching for secondary signal components. 
	  The light curves have been shifted vertically for better visibility. 
	  The lowermost light curve resulted from the subtraction of the large 
	  amplitude pulsation component (including the $14$-th harmonics) from 
	  the systematics-filtered TFA(FOUR) light curve.}
\label{lc_c03}
\end{figure}

%
%
\begin{figure}
 \vspace{0pt}
 \centering
 \includegraphics[angle=-0,width=75mm]{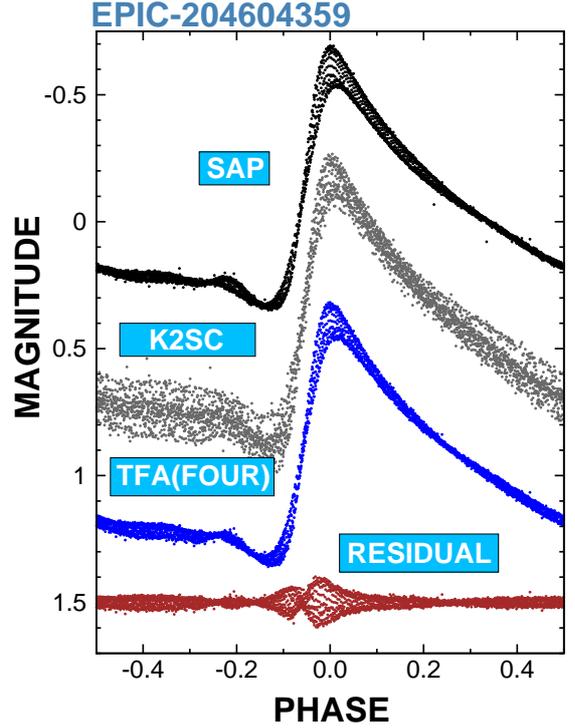}
 \caption{An example from Campaign 2 on the performance of our method in 
          the case of low level of systematics. For comparison, the signal 
	  processing method based on a very flexible Gaussian Process model 
	  (K2SC, Agrain et al.~\cite{aigrain2016}), is also shown.}
\label{lc_c02}
\end{figure}

%
%
\begin{figure}
 \vspace{0pt}
 \centering
 \includegraphics[angle=-0,width=75mm]{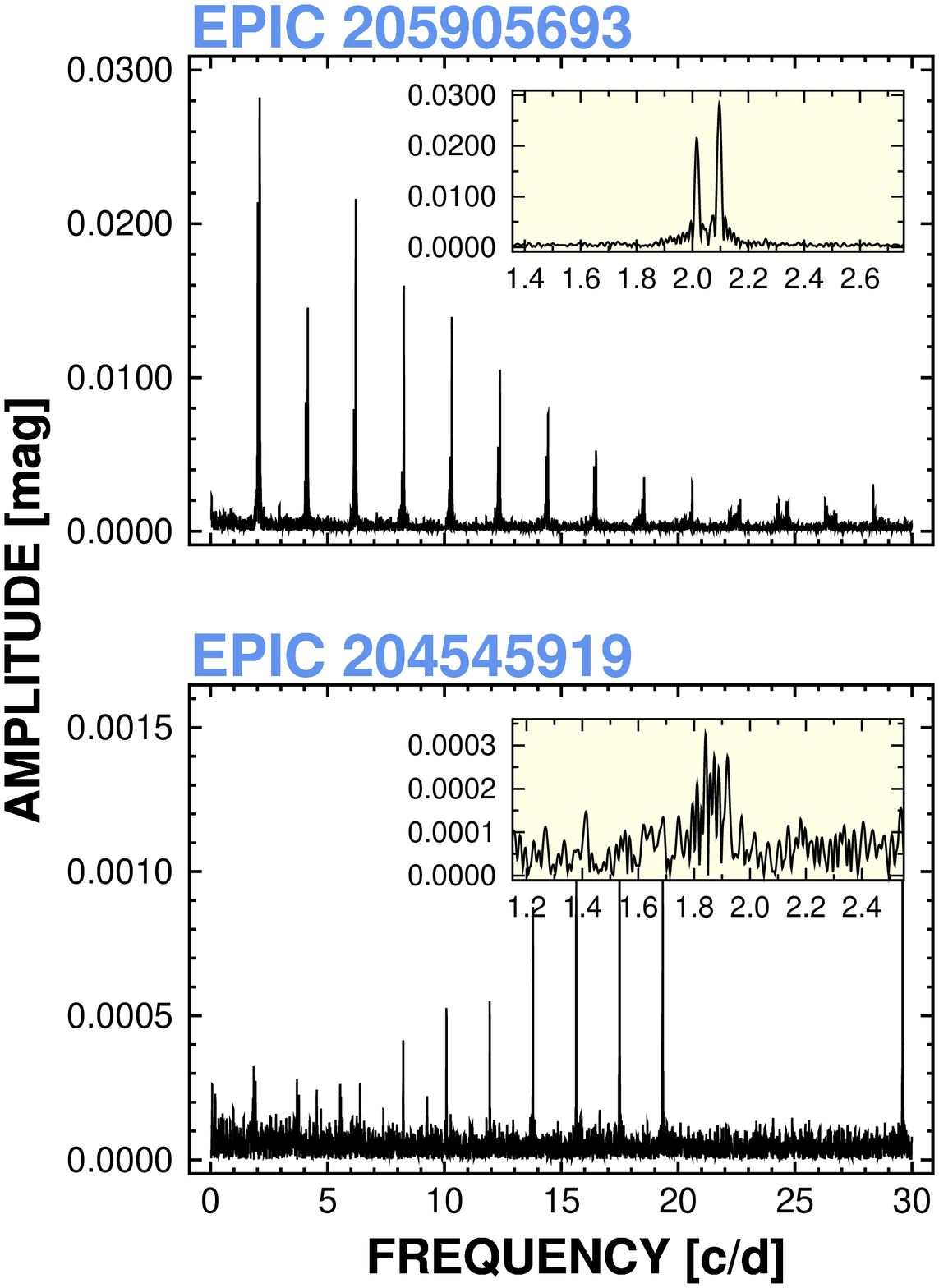}
 \caption{Examples for high (upper panel) and low SNR (lower panel) 
          detections. The amplitude spectra shown are the results of 
	  the DFT analysis of the TFA(FOUR) time series by using the 
	  the fundamental pulsation frequencies and their first $14$ 
	  harmonics both in the TFA filtering and in the pre-whitening. 
	  The peaks come either from the BL side lobes or from the 
	  high-frequency components leaked through the Nyquist frequency 
	  at $\sim 24.47$~c/d.}
\label{lc_c03_c02}
\end{figure}

We show two examples on the performance of the method used in this paper. 
Fig.~\ref{lc_c03} shows a case when the SAP data suffer frome some well 
visible systematics. Our focus of interest is the light curve labelled by 
TFA(FOUR). This light curve obtained after subtracting the best combination 
of the co-trending light curves from the SAP data (i.e., 
TFA(FOUR)$\rightarrow T(i) - a_0 - \sum_{j=1}^{N_{\rm TFA}}a_{\rm j} U_{\rm j}(i)$, 
see Sect.~2). In principle, these data should contain only the modulated 
pulsation of the star. Indeed, the filtered light curve shows a very clean 
pattern of modulation. In comparison, it is also visible that although the 
standard {\em Kepler} pipeline (PDC-SAP, see Smith et al.~\cite{smith2012}) 
helps to alleviate signal distortion at some degree, there are obvious 
remnants of systematics, leaving a less cleanly defined BL modulation on 
the filtered light curve. 

In another example in Fig.~\ref{lc_c02} we show the rare case when the 
systematics are small. Here we expect the method leaving the SAP signal 
basically intact. A brief visual inspection shows that it is indeed the 
case for the TFA(FOUR) light curve, but the filtering method employing 
a Gaussian Process model (K2SC, see Agrain et al.~\cite{aigrain2016}) 
apparently introduces additional noise in the data. These two examples 
show that employing a model best suited to the time series under 
scrutiny -- in this case a periodic signal with a Fourier representation -- 
greatly helps in the separation of systematics from the physically 
relevant signal content. Yet another important difference between our 
method and both of those mentioned above, is the high number of 
co-trending time series used by TFA. In this way we can handle a larger 
variety of systematics than relying only the leading PCA components, 
as in the case of the PDC pipeline.   

The detection of the BL modulation is illustrated in Fig.~\ref{lc_c03_c02}. 
The upper panel shows the ``easy'' case: well-separated modulation side 
lobes with high amplitudes. The lower panel shows one of the cases with 
jammed power content at the fundamental frequency. In these cases either 
the low modulation amplitude or the inherently complex nature of the 
modulation (or both) prevent a reliable estimation of the modulation 
period. Nevertheless, we classified also these cases as BL, since the 
significant power content at the fundamental frequency.     

With multiple visual inspection of the frequency spectra of the $237$ 
stars, we ended up with the following criterion for classifying a RRab 
stars as a BL star. {\em If at any stage of the pre-whitening cycle an 
excess of power occurs with $SNR > 6$ at a frequency $f$, 
satisfying $|f-f_0|<0.1$~c/d, and this feature proves to be stable 
against changing the number of co-trending time series, then the star 
is classified as a BL star}. It is important to note that only about a 
dozen stars required deeper examinations in respect of this criterion. 
Once the side lobe/lobes is/are identified, two parameters were attached 
to the star to characterize its BL properties: a) maximum side lobe 
amplitude, $A_{\rm m}=MAX\{A_{\rm i}; i=1,2, ..., N_{\rm s}\}$, where 
$\{A_{\rm i}\}$ are the side lobe amplitudes (altogether $N_{\rm s}$ 
of them), and b) the modulation frequency $f_{\rm m}=f_0-f(A_{\rm m})$. 
Please note that we selected these two parameters for the rough 
characterization of the BL stars, mainly to establish the distribution 
of $A_{\rm m}$, which is directly connected to the issue of detectability, 
the focus of this work. 

\begin{figure}
 \vspace{0pt}
 \centering
 \includegraphics[angle=-90,width=85mm]{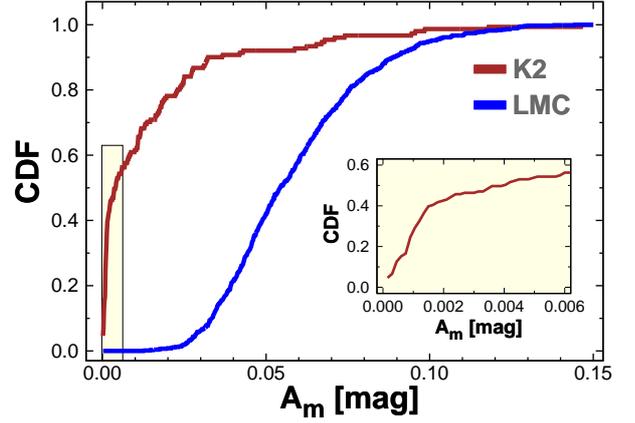}
 \caption{Cumulative distribution function of the maximum side lobe 
          amplitudes of the $138$ BL stars identified in this work. 
	  For comparison, the same function is also shown for the $731$ 
	  BL stars identified in the MACHO LMC sample by 
	  Alcock et al.~(\cite{alcock2003}). The inset shows the blow-up 
	  of the region indicated in the main plot.}
\label{cdf_am}
\end{figure}

\begin{figure}
 \vspace{0pt}
 \centering
 \includegraphics[angle=-90,width=85mm]{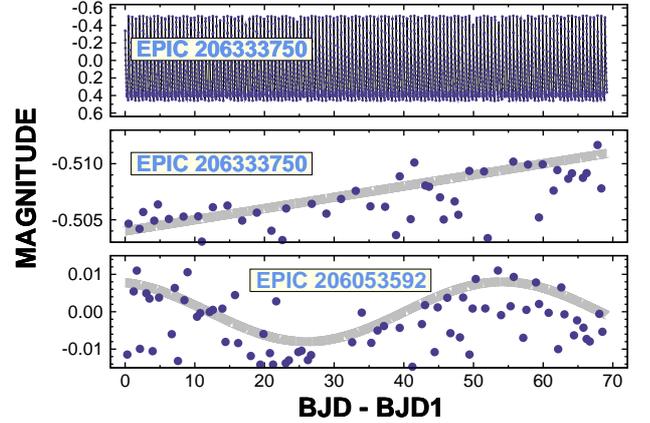}
 \caption{Examples for small amplitude modulations from C03. 
          {\em Uppermost panel}: light curve plotted on the scale 
	  of the full range of variation. {\em Middle panel}: 
	  zoom on the upper envelope. {\em Lower panel}: as in 
	  the middle panel, but for a star with a modulation 
	  period of $56$~days. The gray lines are only for guiding 
	  the eyes in tracing the amplitude variation.}
\label{low_am}
\end{figure}

As we have already mentioned, we found only $151$ stars from the original 
set of $237$ stars that can be classified as RRab stars. With the criterion 
mentioned, $138$ of them proved to be BL stars. This implies $91$\% 
observed occurrence rate, which is the highest value reported so far. 
The cumulative distribution function of the observed maximum side lobe 
amplitudes is shown in Fig.~\ref{cdf_am}. We see that some $40$\% of the 
BL population originates from stars showing a rather small modulation,  
under $0.002$~mag.\footnote{It is recalled that this is the size of the 
largest side lobe. Because additional components are frequently present, 
the total amplitude of the modulation is, in general, larger.} It is also 
observed that the CDF has a break at $\sim 0.0015$~mag, suggesting perhaps 
two different populations of BL stars (or maybe even three, if we consider 
the further break suspectible at the large $\{A_{\rm m}\}$ limit of 
$0.03$~mag). Because of the importance of the population with low 
$\{A_{\rm m}\}$, in Fig.~\ref{low_am} we illustrate the actual observable 
light curve variations for two cases. Partially because of the long cadence 
of the data, in both cases the amplitude variations are not visible in 
the full-scale light curves. Therefore, we zoomed to the upper envelopes 
in the middle and lower panels.   

%
%
  
%
%

%
%
\section{The underlying rate}
It is clear that the orders of magnitude difference between the accuracy 
of the MACHO and K2 data should be an important source of the stunning 
difference in the observed $\{A_{\rm m}\}$ distributions. Here we test 
the effect of the observational noise on the derived occurrence rates. 

Our testbase for the K2 data is the set of $151$ RRab stars analyzed 
in this paper. For the MACHO data we use a larger set of non-BL (single 
period) RRab stars consisting of some $1700$ stars from our earlier 
analysis in Alcock et al.~(\cite{alcock2003}).\footnote{We aim at a 
larger statistical sample, therefore we employ the subset of the non-BL 
stars. By using the BL stars, we get the same result.}  
The tests constitutes of the following steps. First, with the aid of an 
{\em injected signal}\footnote{Importantly, we inject the test signal in 
the SAP data and proceed as with the real (observed) data. Injecting 
the test signal in the TFA-filtered data would falsify (i.e., increase) 
our detection rates, because of the considerable lower noise resulting 
from the filtering.} analysis, we check the detection rates at various 
test amplitudes. We use six different amplitudes both for the K2 and 
for the MACHO data and save the result for the occurrence rate test. 
The range of amplitudes ensures that for each star we have at least one 
detection. The frequency of the injected signal is placed at 
$f_{\rm inj}=1.11f_0$. This choice yields a component at a relative 
unpopulated regime for the K2 data, which is important in avoiding 
interference with the signals already present in the data (for the 
MACHO data this is less of a problem, due to the high noise). The 
injected signal is labelled as ``detected'' if the corresponding SNR 
is greater than $6.0$ and the distance of the associated peak from the 
injected value is less than $0.5/T$, where $T$ is the total time span 
of the observations. 

\begin{figure}
 \vspace{0pt}
 \centering
 \includegraphics[angle=-90,width=75mm]{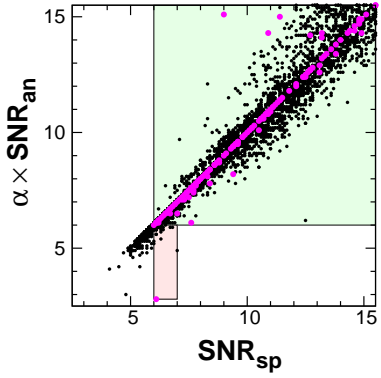}
 \caption{Detected signal-to-noise ratios ($SNR_{\rm sp}$) in the Fourier 
          spectra vs the calibrated analytical signal-to-noise ratio $SNR_{\rm an}$. 
	  Each star is characterized by its own calibration factor $\alpha$ 
	  and $SNR_{\rm an}$. The calibration has been derived from injected 
	  signal tests, as described in the text. The MACHO and K2 stars are 
	  plotted by black and magenta, respectively. Pale green and red rectangles, 
	  respectively, denote the parameter regimes where the two SNR estimates 
	  yield consistent/inconsistent detection result.}
\label{snr_calib}
\end{figure}

Because the estimation of the occurrence rate test requires the knowledge 
of the detection likelihood for any amplitude given by the distribution 
of $\{A_{\rm m}\}$, we calibrate the test $SNR$ values observed in the 
spectra ($SNR_{\rm sp}$) with the analytical form of 
$SNR_{\rm an}=A_{\rm inj}\sqrt{N}/\sigma$ via a simple scaling factor 
$\alpha$.\footnote{Because of the importance of the low-SNR regime, we 
weight heavily (proportionally to $SNR_{\rm sp}^{-6}$) these values 
in the least squares fit for $\alpha$.} Fig.~\ref{snr_calib} shows the 
result of this calibration. We see that except for a negligible small 
fraction of the test values we can tell reliably for each star if the 
injected signal were detectable or not.  

Now we can use the {\em observed} $\{A_{\rm m}\}$ distribution for the K2 
data as shown in Fig.~\ref{cdf_am}, and generate large number of amplitudes 
and check if they are detectable in the stars of the sample. The results 
are summarized in Table~\ref{true_inc}.  

%
%
\begin{table}[!h]
  \caption{Estimates on the observed occurrence rates}
  \label{true_inc}
  \scalebox{1.0}{
  \begin{tabular}{lccc}
  \hline
   Method                        & $R_{\rm true}$  & K2  &  MACHO \\ 
 \hline
 observations                    & -- &  $91\pm2$\% &  \, \, $12\pm0.4$\% \\
 $\{A_{\rm m}\}$ constraint: NO  & $100$\% &  $83\pm3$\% &  $14\pm1$\% \\
 $\{A_{\rm m}\}$ constraint: a   & $100$\% &  $85\pm3$\% &  $13\pm1$\% \\
 $\{A_{\rm m}\}$ constraint: b   & $100$\% &  $91\pm2$\% &  $14\pm1$\% \\
 $\{A_{\rm m}\}$ constraint: NO  & $90$\%  &  $74\pm3$\% &  $13\pm1$\% \\
 $\{A_{\rm m}\}$ constraint: a   & $90$\%  &  $76\pm3$\% &  $12\pm1$\% \\
 $\{A_{\rm m}\}$ constraint: b   & $90$\%  &  $81\pm3$\% &  $12\pm1$\% \\
\hline
\end{tabular}}
\begin{flushleft}
{\bf Notes:}\\
\vbox{- All tests have been made under the assumption that the true 
distribution of the maximum modulation side lobes $\{A_{\rm m}\}$ 
follow the observed distribution as derived on the K2 data (see 
Fig.~\ref{cdf_am}).\\ 
- The type of constraint employed in the distribution of $\{A_{\rm m}\}$ 
is given by the letters {\em a} and {\em b}, corresponding to the 
regions indicated in Fig.~\ref{p_log_am}.\\
- Statistical errors attached to the observed values are based on 
Alcock et al.~(\cite{alcock2003}).} 
\end{flushleft}
\end{table}
%

%
%

\begin{figure}
 \vspace{0pt}
 \centering
 \includegraphics[angle=-90,width=85mm]{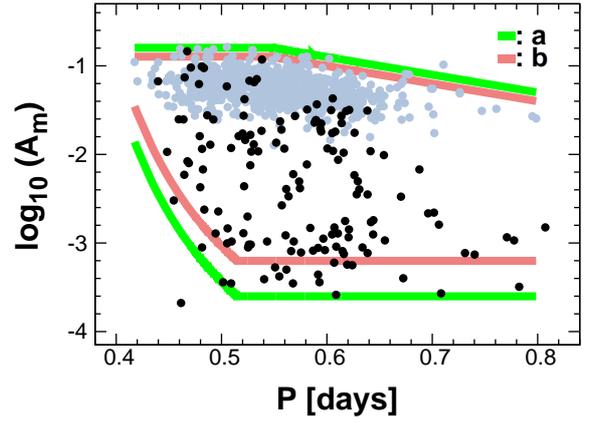}
 \caption{Observed modulation amplitudes (Fourier amplitudes -- in [mag] 
          -- of the largest sidelobes in the vicinity of the frequency 
	  of the fundamental mode) versus pulsation period. Black dots 
	  are for the K2 sample of this work, whereas pale ones are for 
	  the Blazhko stars from the MACHO survey of 
	  Alcock et al.~(\cite{alcock2003}). The various bounds of the 
	  region denoted by {\em a} and {\em b} are used in the estimation 
	  of the underlying occurrence rates.}
\label{p_log_am}
\end{figure}

In the first row we show the observed rates. In the next three rows we assume 
$100$\% true occurrence rate, and impose various constraints on the injected 
signal amplitude. In the simplest case we assume that each star has equal 
probability to sample the full range of modulation amplitudes according to the 
observed distribution on the K2 dataset. This assumption leads to a significant 
underestimation of the observed K2 rate. An obvious guess for a possible 
source of the underestimation is that the (unknown) dependence of the BL 
phenomenon on the stellar parameters cuts out low amplitude modulations 
for specific parameter combinations. We tested this possibility by examining 
the period dependence of the modulation amplitudes. Fig.~\ref{p_log_am} shows 
that indeed, at short periods, we have an apparent lack of BL stars. Applying 
the constraints imposed by the lines labelled by {\em a}, we get an increase 
(see third row), but it is still insufficient to reproduce the observed rate 
for the K2 data. Shrinking further the allowed region (case {\em b}) we arrive 
to a nice match for the K2 data. We note that at this level we have no 
knowledge on the actual parameter space of the BL phenomenon. It might be 
more complicated than the one used here (i.e., a simple smooth amplitude cut). 
Therefore, the deeper cut seen in case {\em b} could be less severe as it 
might seem from the several outliers in Fig.~\ref{p_log_am} for this 
particular cut. 

Because of the higher noise level of the MACHO data, they yield roughly the 
same, slightly discrepant value, independently of the restrictions posed on 
the modulation amplitude. It seems that the only way to reach the observed 
value is to decrease the underlying occurrence rate. Indeed, as shown 
in the last three rows, assuming $90$\% true rate yields the desired rate 
for the MACHO/LMC data. Unfortunately, the lower rate completely spoils the 
fine agreement for the K2 data. This may imply some real difference in the 
BL occurrence rates between the LMC and the Galactic field. 

%
%

%
%
\section{Conclusions}
We analyzed $151$ RR~Lyrae stars from the K2 fields C01--C04, classified 
earlier by Armstrong et al.~(\cite{armstrong2016}) as fundamental mode 
(RRab) stars and searched for Blazhko variability, generally characterized 
as stars displaying close frequency components to that of the fundamental 
mode. Our analysis adapts the method used widespread in the field of 
transiting extrasolar planets to filter out systematics from ground- and 
space-based data. These systematics act as colored noise in the frequency 
analysis, and falsify the result for signals with amplitudes similar to 
those of the systematics. The modified method takes into consideration of 
the large amplitude component (stellar pulsation) of the RRab signal and 
maintains its shape with a simultaneous systematics filtering. Our analysis 
has led to the following results. 
\begin{itemize}
\item
We report an occurrence rate of $91$\% for the Blazhko phenomenon among the 
stars analyzed. This is the highest rate detected so far.
\item
After correcting for observational bias due to finite observational accuracy, 
we find that the underlying (true) rate should be very close to $100$\% in 
the part of the Galactic field covered by C01--C04.
\item
This high true rate is confirmed by a similar debiasing applied to the 
MACHO data on LMC (Alcock et al.~\cite{alcock2003}), although the optimum 
rate turned out to be closer to $90$\% for this dataset. 
\item
The cumulative distribution function of the modulation side lobe amplitudes 
has a clear break at $0.0015$~mag, leading to a conjecture on the possibility 
of the existence of a different subclass of Blazhko stars at small modulation 
amplitudes. 
\item
Because of the very common nature of the Blazhko phenomenon, it seems likely 
that ideas employing radial mode resonances will face difficulties in explaining 
this level of commonality. 
\end{itemize}

%
\begin{acknowledgements}
I thank to David Armstrong for a quick reply to my inquiry on the data availability. 
I also appreciate the information given on the analysis of the WASP data by 
Paul Greer. Special thanks are due to Eric Petigura for making the K2 time 
series data available to the community in simple format through the IPAC ExoFOP 
site. This research has made use of the NASA Exoplanet Archive, which is 
operated by the California Institute of Technology, under contract with the 
National Aeronautics and Space Administration under the Exoplanet Exploration 
Program.  
\end{acknowledgements}

\bibliographystyle{aa} 

\end{document}